\documentclass{kapproc} 

\usepackage{procps}

\usepackage[dvips]{graphicx}

\upperandlowercase

\kluwerbib 

\begin{document}
\articletitle[Barred Galaxies] {How Barred is the NIR Nearby Universe? An analysis using 2MASS}

\author{Karin Menendez-Delmestre, Kartik Sheth, Nick Scoville, Tom
  Jarett\altaffilmark{1},  Eva Schinnerer\altaffilmark{2},Michael
  W. Regan\altaffilmark{3}, and David Block\altaffilmark{4}}
 
\affil{\altaffilmark{1}California Institute of Technology, Pasadena,
  CA, \altaffilmark{2}National Radio Astronomy Observatory,
  \altaffilmark{3}Space Telescope Science Institute,
  \altaffilmark{4}University of the Witwatersrand }

\begin{abstract}
We determine a firm lower limit to the bar fraction of 0.58 in the nearby
universe using J+H+K-band images for 134
spirals from 2MASS. With a mean 
deprojected semi-major axis of  5.1 kpc, and a mean deprojected 
ellipticity of 0.45 this  local bar sample lays the ground work for 
studies on bar formation and evolution at 
high redshift.  
\end{abstract}

The Two Micron All Sky Survey (2MASS) offers a rich source of images
for nearby galaxies in K-band, the optimal band for detecting bars.
We selected all spiral galaxies (S0/a--Sd) with $i < 65^{\circ}$,
which resulted in a final sample size of 134 spirals.  We
analysed each galaxy using ellipse-fitting to detect
the presence of bars (Fig. 1) and characterise their properties.  Our
method consists of two signature criteria ($\bigtriangleup \epsilon
\geq 0.1$, $\bigtriangleup PA \geq 10^{\circ}$ ) that can be applied
to any sample.  We find a firm lower limit to the bar fraction of
0.58, which is not significantly different from the fraction inferred
from B-band observations (de Vaucouleurs, RC3).  This indicates that
bar morphology is still recognizable in B-band.  However, bar
properties are better traced in K-band, where star formation and dust
obscuration are less severe.  We found that bars occupy in average
$\sim$35\% of the galaxy disk, with a mean bar axial ratio very close
to 2:1 (Fig. 2).  A marked dearth of strong bars is in agreement with
previous studies (Buta \& Block, 2001) and has been interpreted as an
indication that bars at the current epoch are a second, third, or
later generation (Bournaud \& Combes 2002).  A slight trend between
the bar size and the galaxy size/brightness is also found: larger bars
appear to live in larger and more massive galaxies.  However, no
similar trend with the bar strength was found.  For more details on
the method and analysis, refer to Sheth et al in this volume.

Our study of bar length and ellipticity sets the stage for the
evolution of these properties as a function of redshift.  Results from
high redshift studies (e.g., Sheth et al. 2003) may then be compared
to this nearby sample in order to put more stringent constraints on
bar evolution and, ultimately, galaxy evolution.

\begin{figure}[ht]
\vskip -.5in
\centerline{\includegraphics[height=2in]{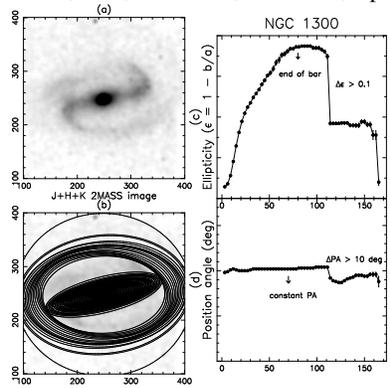}}
\caption{(a) ngc1300 (b) with fitted ellipses. (c) Bar signature:  monotonic increase in ellipticity and (d) constant PA followed by  sharp change in both parameters.}
\end{figure}


\begin{figure}[ht]
\centerline{\includegraphics[height=3.65in]{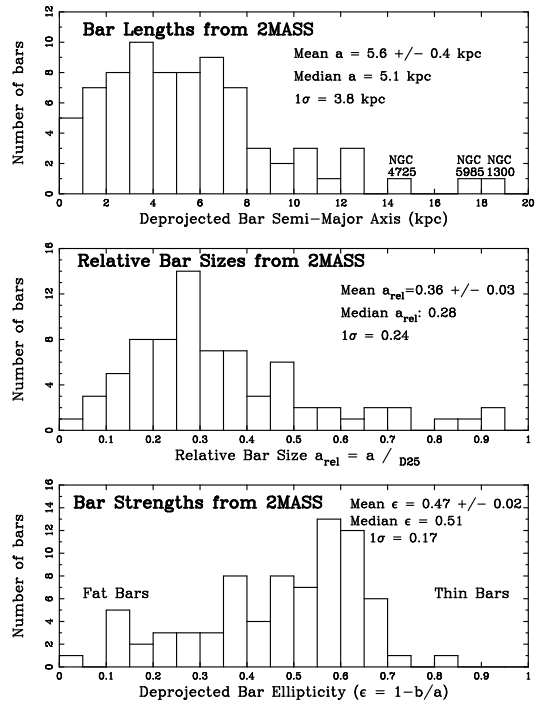}}
\caption{Histograms of bar lengths, relative size and bar strenghts}

\end{figure}





\end{document}